\documentclass[twocolumn,aps,english,prl,showpacs]{revtex4-1}

\usepackage{amsmath}
\usepackage{graphicx}
\usepackage{amssymb}
\usepackage{amsbsy}

\newcommand{\tiem}[1]{{\text{\tiny{#1}}}}

\makeatletter

\newcommand{\lyxmathsym}[1]{\ifmmode\begingroup\def\b@ld{bold}
  \text{\ifx\math@version\b@ld\bfseries\fi#1}\endgroup\else#1\fi}

\@ifundefined{textcolor}{}
{%
 \definecolor{BLACK}{gray}{0}
 \definecolor{WHITE}{gray}{1}
 \definecolor{RED}{rgb}{1,0,0}
 \definecolor{GREEN}{rgb}{0,1,0}
 \definecolor{BLUE}{rgb}{0,0,1}
 \definecolor{CYAN}{cmyk}{1,0,0,0}
 \definecolor{MAGENTA}{cmyk}{0,1,0,0}
 \definecolor{YELLOW}{cmyk}{0,0,1,0}
 }

\@ifundefined{definecolor}
 {\@ifundefined{definecolor}
 {\usepackage{color}}{}
}{}

\usepackage{relsize}

\makeatletter
\DeclareRobustCommand{\lyxmathsym}[1]{\ifmmode\begingroup\def\b@ld{bold}
  \def\rmorbf##1{\ifx\math@version\b@ld\textbf{##1}\else\textrm{##1}\fi}
  \mathchoice{\hbox{\rmorbf{#1}}}{\hbox{\rmorbf{#1}}}
  {\hbox{\smaller[2]\rmorbf{#1}}}{\hbox{\smaller[3]\rmorbf{#1}}}
  \endgroup\else#1\fi}

\makeatother

\begin{document}

\title{Ultrafast dynamical path for the switching of a ferrimagnet after femtosecond heating}

\author{U. Atxitia$^{1}$}

\author{T. Ostler$^{1}$}

\author{J. Barker$^{1}$}

\author{R. F. L. Evans$^{1}$}

\author{R. W. Chantrell$^{1}$}

\author{O. Chubykalo-Fesenko$^{2}$}

\affiliation{$^{1}$Department of Physics, University of York,	Heslington, York YO10 5DD United Kingdom}
\affiliation{$^{2}$Instituto de Ciencia de Materiales de Madrid, CSIC, Cantoblanco, 28049 Madrid, Spain}

\date{\today }

\begin{abstract}
Ultrafast laser-induced magnetic switching in rare earth-transition metal ferrimagnetic alloys has recently been reported to occur
by ultrafast heating alone.
Using atomistic simulations and a ferrimagnetic Landau-Lifshitz-Bloch formalism, we demonstrate that for
 switching to occur it is necessary that angular momentum is transferred from the longitudinal to transverse magnetization
components in the transition metal.
This dynamical path leads to the transfer of the angular momentum to the rare earth metal and magnetization switching with subsequent ultrafast precession caused by the inter-sublattice exchange field
on the atomic scale.
\end{abstract}

 \maketitle

The behavior of magnetization dynamics triggered by an ultrafast laser stimulus is a topic of intense research
interest in both fundamental and applied magnetism~\cite{Siegmann}.
A range of studies using ultrafast laser pulses have shown very different timescales of demagnetization for different
materials; from 100 fs  in Ni~\cite{BeaurepairePRL1996} to  100 ps in Gd~\cite{WeistrukPRL2011}.
Any potential applications utilizing such a mechanism would require,
not only  ultrafast demagnetization, but also  controlled magnetization switching.

Magnetization reversal induced by an ultrafast laser pulse has been reported in the ferrimagnet GdFeCo,
together with a rich variety of phenomena \cite{StanciuPRL2007,Hansteen,VahaplarPRL2009,Radu2011,Ostler2012}.
  Several hypotheses have been put forward to explain the observed magnetization switching: crossing of the
angular momentum compensation point~\cite{StanciuPRL2007}, the Inverse Faraday Effect~\cite{Hansteen}, and
its combination with ultrafast heating~\cite{VahaplarPRL2009}.
It has been shown that the rare earth (RE) responds more slowly to the laser pulse than the transition
metal (TM)~\cite{Radu2011}, even though the sublattices are strongly exchange coupled. Intriguingly, Radu
\emph{et al.} \cite{Radu2011}
show experimentally and theoretically the existence of a transient ferromagnetic-like state,
whereby the two sublattices align against their exchange interaction, existing for
a few hundred femtoseconds.
Recently \cite{Ostler2012}, the atomistic model outlined in \cite{Radu2011,Ostler2012} predicted the phenomenon of magnetization
reversal induced by heat alone, in the absence of any external field; a prediction verified experimentally.
This remarkable result opens many interesting possibilities in terms of ultrafast magnetization reversal and
potential areas of practical exploitation, however a complete theoretical understanding of this effect is currently missing.

In magnets consisting of more than one magnetic species, excitation of the spins on a time scale comparable with that of the
inter-sublattice exchange takes the sublattices out of equilibrium with each other. It is in this regime where the thermally
driven switching of ferrimagnetic GdFeCo occurs. A recent study by Mentink \emph{et al.} \cite{Mentalnik2012} proposed an explanation of
the process using a phenomenological model of the magnetization dynamics, which assumes the additive character of two relaxation mechanisms:
one governed by the inter-sublattice exchange and another by the relativistic contribution (coupling to external degrees of freedom).  The model is based on
the physically plausible argument that the switching is driven by angular momentum transfer in the exchange-dominated regime.
  However, the assumption of a linear path to reversal allows the angular momentum transfer to occur through longitudinal components only,
since the perpendicular components are neglected.
Additionally, the dynamical equation in Ref. \cite{Mentalnik2012} was derived from the Onsager principle,
generally valid for small deviations from the equilibrium only.  Thus far,
 a complete explanation  of the heat driven, ultrafast reversal process remains illusive.

In this Letter we demonstrate that the switching of magnetization in a ferrimagnet after femtosecond heating
is due to the transfer of angular momentum from the longitudinal to the transverse magnetization components in
the TM and consequent transfer of the angular momentum through perpendicular components to the RE.
We present a general formalism, leading to  a macroscopic dynamical equation for a ferrimagnet.
This is in the form of a Landau-Lifshitz-Bloch (LLB) equation,
in which, unlike the phenomenological model of Ref. \onlinecite{Mentalnik2012}, the two relaxation mechanisms are not additive.
Our theory  gives the non-equilibrium conditions necessary for
this angular momentum transfer to happen and thus to produce the precessional rather than linear reversal
as suggested in Ref. \onlinecite{Mentalnik2012}.
These predictions are supported by calculations using an atomistic model based on the Heisenberg exchange Hamiltonian
with Langevin dynamics.

In the absence of any external stimulus, the energetics of the atomistic spin model are described purely by exchange interactions,
given by the spin Hamiltonian:
\begin{equation}
\mathcal{H}= -\sum_{j<i} J_{ij} \mathbf{S}_{i} \cdot \mathbf{S}_{j}
\label{eq:fullham}
\end{equation}
where $J_{ij}$ is the  exchange integral between  spins $i$ and $j$ ($i, j$ are lattice sites), and where $j$ runs over first nearest neighbors only, $\mathbf{S}_{i}$
is the normalized magnetic moment $\left|\mathbf{S}_{i}\right|=1$. We model the magnetization dynamics of the system
using the Landau-Lifshitz-Gilbert (LLG) equation with Langevin dynamics, as detailed in Ref.~\onlinecite{Ostler2011}.
The system consists of
$\mathcal{N} \times\mathcal{N} \times \mathcal{N}$  cells in a fcc structure lattice which we populate  with a random distribution of TM
and RE ions in the desired concentration $q$ and $x=1-q$.
To simulate the effect of an ultrafast heat pulse we use a step-like temperature pulse of duration 500 fs with a  value  of
$T=T_{\text{max}}$.
The model predicts the switching of GdFeCo compound under the ultrafast heat alone, as demonstrated in Ref.~\onlinecite{Ostler2012}.
Atomistic models have proven to be a powerful tool in predicting heat-induced switching,
but fail to provide a simple picture for the cause of its physical origin.

However, the macroscopic LLB equation has been demonstrated to be an adequate approach, allowing
a simple description of ultrafast magnetization phenomena \cite{Atxitia2010, Sultan},
 but up to now it existed only for a single species ferromagnet \cite{Garanin}.
Recently \cite{LLBferri}, we have derived the LLB equation for a two species system
which describes the average magnetization dynamics in each sublattice
$\mathbf{m}_\nu=\left\langle\mathbf{S}^{\nu}_i\right\rangle$,
where $\nu$ stands for TM or RE sublattice in this case and $i$ for spins in the sublattice $\nu$.
Importantly, unlike the approach used in Ref. \cite{Mentalnik2012}, the derivation does not use the Onsager principle
and is thus valid far from equilibrium.
\begin{figure}[tb!]
\includegraphics[width=8.0cm, trim=10 20 10 0]{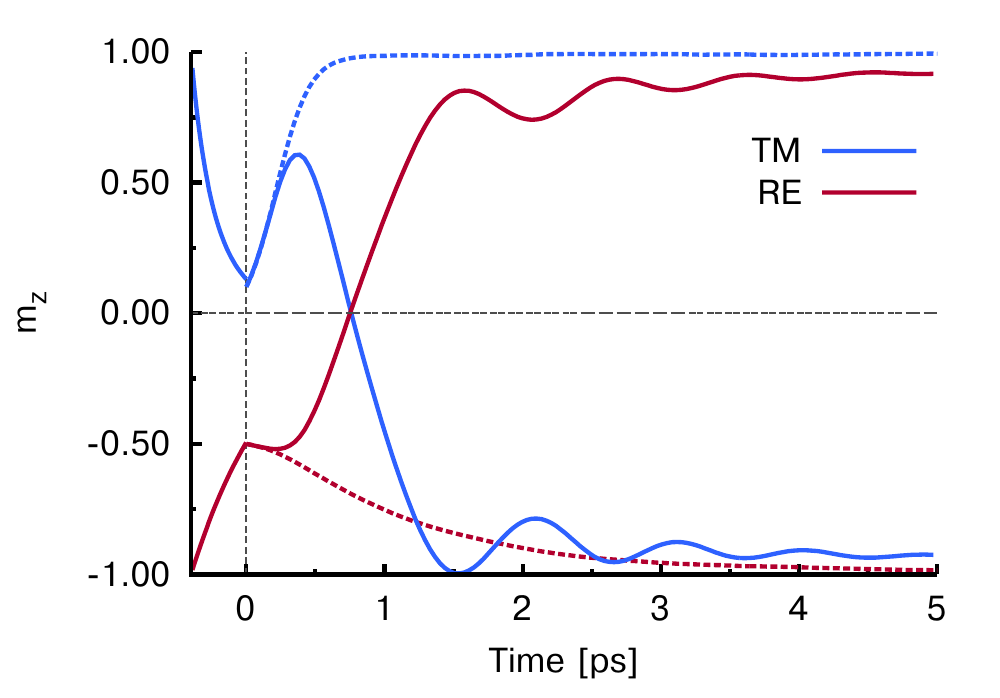}
\caption{Numerical integration of the switching behavior for the non-stochastic LLB with a
small angle (15 degrees) between sublattices (solid lines). Without the angle (dashed lines) switching does not occur,
as predicted. The time $t=0$ corresponds to the end of the square shaped laser pulse with $T_{\text{max}}=1500$ K.
For the integration at temperatures above $T_C$ we use the paramagnetic version of the ferrimagnetic LLB equation with MFA \cite{LLBferri}.}
\label{fig:LLB-switching}
\end{figure}

In the absence of an applied field and anisotropy, the LLB equation for the TM is written as:
\begin{equation}
\label{eq:LLBT}
\frac{1}{|\gamma_{\tiem{T}}|}\frac{\mathrm{d} \mathbf{m}_{\tiem{T}}}{\mathrm{d} t}{=}
{-}\mathbf{m}_{\tiem{T}}{\times} \Big[ \mathbf{H}^{\tiem{EX}}_{\tiem{T}}{+}
\frac{\alpha_{\tiem{T}}^{\perp}}{m^2_{\tiem{T}}}\mathbf{m}_{\tiem{T}} {\times} \mathbf{H}^{\tiem{EX}}_{\tiem{T}} \Big]
{+}\alpha_{\tiem{T}}^{\|} H^{\|}_{\tiem{T}}
\mathbf{m}_{\tiem{T}}\textbf{,}
\end{equation}
with a complementary equation for the RE.
The exchange field from the RE is calculated via the mean-field approximation (MFA) of the impurity model
presented in \cite{Ostler2011} as $\mathbf{H}^{\tiem{EX}}_{\tiem{T}}=(J_{0,\tiem{TR}}/\mu_{\tiem{T}})\mathbf{m}_{\tiem{R}}$,
where $J_{0,\tiem{TR}}=xzJ_{\tiem{TR}}$,  $x$ is the impurity content,
$z$ the number of nearest TM neighbors in the ordered lattice and
$J_{\tiem{TR}}<0$  the inter-sublattice exchange parameter.
The TM magnetic moment is  denoted  $\mu_{T}$,  $\gamma_{\tiem{T}}$
is the gyromagnetic ratio for the TM lattice, $\alpha_{\tiem{T}}^{\|}(T)$ and $\alpha_{\tiem{T}}^{\bot}(T)$ are
temperature-dependent TM longitudinal and transverse damping parameters,  linearly proportional to
the intrinsic coupling to the bath parameter $\lambda_{\tiem{T}}$  \cite{LLBferri}.
The longitudinal effective field in Eq. \eqref{eq:LLBT} reads
\begin{equation}
  H^{\|}_{\tiem{T}}=
\frac{\Gamma_{\tiem{TT}}}{2}
\left(1-\frac{m_{\tiem{T}}^{2}}
{m_{e,\tiem{T}}^{2}}\right)-\frac{\Gamma_{\tiem{TR}}}
{2}
\left(1-\frac{\tau_{\tiem{R}}^{2}}
{\tau_{e,\tiem{R}}^{2}}\right)
\textbf{,}
\label{eq:longitudinalrates}
\end{equation}
where $\tau_{\tiem{R}}=|(\mathbf{m}_{\tiem{T}}\cdot\mathbf{m}_{\tiem{R}})|/m_{\tiem{T}}$ is the absolute value of
the projection of the RE magnetization onto the TM magnetization and $\tau_{e,\tiem{R}}$ is its equilibrium value.
The rate parameters in Eq. \eqref{eq:longitudinalrates} read
\begin{eqnarray}
 \Gamma_{\tiem{TT}}=\frac{1}{\widetilde{\chi}_{\tiem{T},\|}}
\left(1+\frac{|J_{0,\tiem{TR}}|}{\mu_{\tiem{T}}}\widetilde{\chi}_{\tiem{R},\|}
\right), \
\Gamma_{\tiem{TR}}=\frac{|J_{0,\tiem{TR}}|}{\mu_{\tiem{T}}} \frac{\tau_{e,\tiem{R}}}{m_{e,\tiem{T}}}.
\label{Rates}
\end{eqnarray}
They are temperature-dependent  via the equilibrium magnetizations and partial longitudinal susceptibilities
$\widetilde{\chi}_{\tiem{T},\|}=\left( \partial m_{\tiem{T}}/ \partial H \right)_{H \rightarrow 0}$,
$\widetilde{\chi}_{\tiem{R},\|}=\left( \partial m_{\tiem{R}}/ \partial H \right)_{H \rightarrow 0}$, evaluated in the MFA
in the presence of inter-sublattice  and
intra-sublattice exchange \cite{LLBferri}.

In Eq. \eqref{eq:LLBT} the first term in the r.h.s. describes the precession of
the TM magnetization, $\mathbf{m}_{\tiem{T}}$, around the exchange field
produced by the RE sublattice.  Although this term
conserves the magnetization modulus, $m_{\tiem{T}}$, it  allows transfer of angular momentum
between lattices.
 The second term in Eq. \eqref{eq:LLBT} describes the relaxation of $\mathbf{m}_{\tiem{T}}$ towards
 the antiparallel alignment between both sublattice  magnetizations.
Finally, the third  term in Eq. \eqref{eq:LLBT} defines the longitudinal relaxation, comprised of;
the difference between
relaxation coming from the deviations of TM magnetization from  equilibrium and those of RE.
In the ferrimagnetic LLB all three terms  act on the timescale given by the exchange interactions  in comparison
to the ferromagnetic LLB case, where the longitudinal and transverse motion have very different timescales \cite{Chubykalo,AtxitiaQ}.

\begin{figure}[!tb]
\includegraphics[width=9cm]{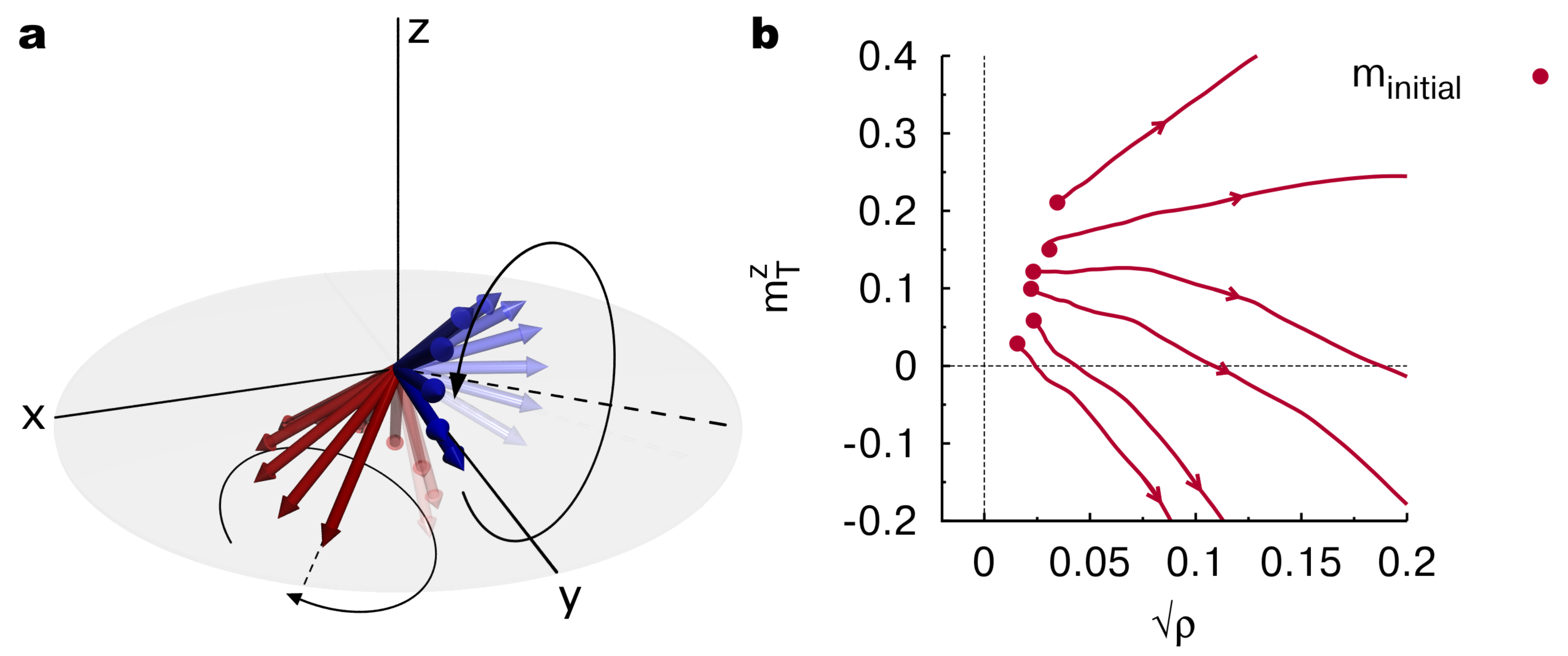}
\caption{ (a)  Precession of sublattice magnetizations around the exchange field of each other in the macroscopic (LLB) description.
After the action of an ultrafast laser pulse
 the large amplitude of the TM precession causes it to cross $m_z = 0$,
and for sufficiently large angular momentum transfer,
the angle between sublattices becomes small.
After cooling the dominance of the TM sublattice forces the RE to realign along the opposite direction,
completing the switching process.
(b)  Trajectories of the parallel and transverse magnetization components
for TM calculated via atomistic simulations of the Heisenberg model (\ref{eq:fullham}) at different maximum pulse temperatures
$T_{\text{max}}=1000, 1200, 1250, 1300,1350$ and $1400$ K. After the pulse, the temperature is removed, this moment is indicated by small circles. }
\label{schematics}
\end{figure}

Fig.~\ref{fig:LLB-switching} shows the  direct numerical integration of Eq.~\eqref{eq:LLBT}. With initial antiparallel alignment
 of the RE and TM, $\mathbf{m}_{\tiem{T}}\| \ \mathbf{m}_{\tiem{R}}$,  when the temperature is raised 
both sublattice magnetizations are reduced, followed by the linear magnetization recovery path
to the expected ground state [see dashed lines in Fig. \ref{fig:LLB-switching}] and does \emph{not}  produce switching.
 In this case no torque is exerted from one sub-lattice to another as $\mathbf{m}_{\tiem{T}} \times \mathbf{m}_{\tiem{R}}=0$.
However this torque, which allows transfer of angular momentum between sublattices, is always present in
the full atomistic approach with stochastic fields because of the high temperatures reached
during the reversal process.
We can include in Eq. \eqref{eq:LLBT} the presence of this torque  by canting by a small angle  the
two sub-lattices magnetization once the heat pulse is gone or alternatively by the integration of the stochastic
LLB equation~\cite{Evans2012}.
The solid lines in Fig.~\ref{fig:LLB-switching} show the integration of Eq.~\eqref{eq:LLBT} including this angle and shows
reversal.  This small angle generates  a mutual precessional motion which
 occurs due to the exchange field exerted by the opposite sub-lattice and
the transverse relaxation  directed towards the direction of the opposite sub-lattice.
This mutual motion leads to the switching, as illustrated in Fig. \ref{fig:LLB-switching} and is presented schematically in
Fig. \ref{schematics}(a).

Though the longitudinal magnetization
process contributes to the timescale of reversal it does not drive the switching process. Unlike the statement in Ref. \cite{Mentalnik2012},
the longitudinal relaxation  itself cannot change the direction of  $\mathbf{m_{\tiem{T}}}$,
 due to the multiplication of the longitudinal relaxation term in Eq.~\eqref{eq:LLBT}
 by $\mathbf{m_{\tiem{T}}}$. In order to understand the switching
mechanism we therefore need to consider both longitudinal and transverse relaxation.

\begin{figure}[!tb]
\includegraphics[width=8cm, trim=20 20 20 0]{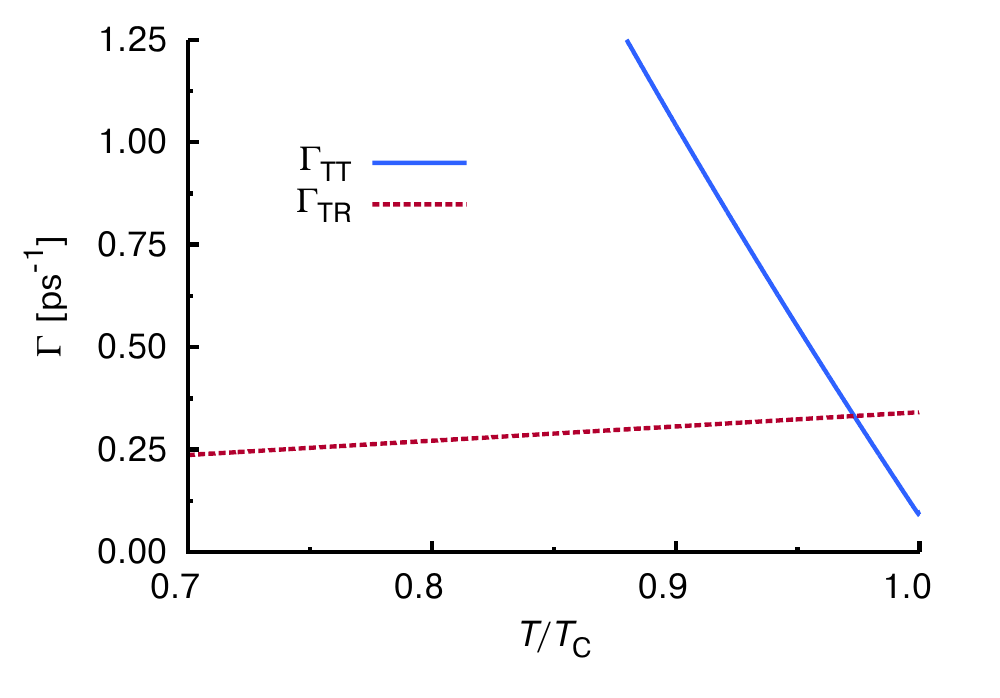}
\caption{Longitudinal relaxation rates as a function of temperature in
the LLB equation, evaluated for GdFeCo parameters.
The dashed line shows the TM-RE relaxation rate and the solid line is that of the TM-TM interaction.
 At low temperatures $\Gamma_{\tiem{TT}} \gg \Gamma_{\tiem{TR}} $, due to the small value of the susceptibility $\widetilde{\chi}_{\tiem{T},\|}$,
therefore the relaxation of the TM magnetization is always to its own equilibrium. However,
at temperatures close to $T_C$, $\Gamma_{\tiem{TT}} < \Gamma_{\tiem{TR}}$,
thus the TM prefers to relax towards the RE magnetization in this regime.}
\label{param}
\end{figure}

Now we demonstrate that at high temperatures the longitudinal relaxation becomes unstable.
This happens because close to $T_C$
the sign of $H^{\|}_{\tiem{T}}$ can change.
In Fig.~\ref{param} we present the temperature dependence of relaxation rates \eqref{Rates}
 evaluated for the parameters of GdFeCo \cite{Ostler2011} in the MFA.
One can see that close to $T_C$: $\Gamma_{\tiem{TT}} < \Gamma_{\tiem{TR}}$.
Firstly  we reduce the LLB equation \eqref{eq:LLBT} to a dynamical system, based on information from atomistic modeling.
We can
assume that slightly before the reversal the initial transverse moments of the sublattices are small (but not zero),
and that the modulus of the TM sublattice is much smaller than that
of the RE ($m_{\tiem{T}}^{z}\ll m_{\tiem{R}}^{0}$), owing to the  faster relaxation time of the TM.
In this approximation  the longitudinal field  is positive $H^{\|}_{\tiem{T}}>0$:
$H^{\|}_{\tiem{T}}\simeq \Gamma_{\tiem{TR}}\frac{m_{\tiem{R}}^{0}}{m_{\tiem{R}}^{e}}$
for the case before the heat pulse is removed ($m_{\tiem{T}} > m_{e,\tiem{T}} = 0$) and
because after the heat pulse is gone the system cools down
$H^{\|}_{\tiem{T}}\simeq [\Gamma_{\tiem{TT}}-\Gamma_{\tiem{TR}}]/2>0$ with $m_{\tiem{T}} \ll m_{e,\tiem{T}}$($T$).
The LLB equation for the TM is reduced to the following system of equations:

\begin{eqnarray}
\frac{\mathrm{d} m_{\tiem{T}}^2}{\mathrm{d} t}&=&2|\gamma_{\tiem{T}}|\alpha_{\tiem{T}}^{\|} H^{\|}_{\tiem{T}}m_{\tiem{T}}^2, \nonumber \\
\frac{\mathrm{d}  \rho}{\mathrm{d} t}&=&-2
\Big[
\alpha_{\tiem{T}}^{\bot}
\Omega_{\tiem{T}}
\sqrt{1-\rho/m_{\tiem{T}}^2} -|\gamma_{\tiem{T}}|\alpha_{\tiem{T}}^{\|} H^{\|}_{\tiem{T}}
\Big]
\rho
\label{eq:rozsystem}
\end{eqnarray}
where $\rho=(m_{\tiem{T}}^{t})^2= (m_{\tiem{T}}^{x})^2+(m_{\tiem{T}}^{y})^2$ is the TM transverse magnetization component,
$\Omega_{\tiem{T}}=m_{\tiem{R}}^{0} |\gamma_{\tiem{T}}| |J_{0,\tiem{TR}}| /\mu_{\tiem{T}}$
is the precessional frequency of the anti-ferromagnetic exchange mode.

The trajectory $\rho=0$ corresponds to a linear dynamical mode.
The standard analysis of the dynamical system \eqref{eq:rozsystem}
shows that for $H^{\|}_{\tiem{T}}>0$ and $m_{\tiem{T}}^z<\alpha_{\tiem{T}}^{\bot}
\Omega_{\tiem{T}}/(|\gamma_{\tiem{T}}|H^{\|}_{\tiem{T}})$
this trajectory becomes unstable.
Before the end of the pulse it is equivalent to $m_{\tiem{T}}>(\alpha_{\tiem{T}}^{\perp}/\alpha_{\tiem{T}}^{\|})m_{e,\tiem{T}}$
which is also easily satisfied, taking into
account that $\alpha_{\tiem{T}}^{\perp}>\alpha_{\tiem{T}}^{\|}$, see Ref. \cite{LLBferri}.
The physical interpretation is that in this case very small perturbations from $\rho=0$ will not be damped but
will lead to the development of a
perpendicular magnetization component, as is indeed observed by the atomistic simulations Fig.~\ref{schematics}(b), in which we use the atomistic model and apply heat pulses of different temperatures to drive the system into different states. The atomistic simulations clearly confirm the development of the perpendicular component.

However, the dynamical system \eqref{eq:rozsystem} alone does not describe the reversal due to the assumption of the static
RE magnetization.  In the same approximation, the LLB equation for the RE reads:
\begin{equation}
\frac{\mathrm{d} m_{\tiem{R}}^{x(y)}}{\mathrm{d} t}=
\pm\Omega_{\tiem{R}}m_{\tiem{T}}^{y(x)}-\frac{\alpha_{\tiem{R}}^{\perp}}{m_{\tiem{R}}^{0}}
\Omega_{\tiem{R}} m_{\tiem{T}}^{x(y)}-|\gamma_{\tiem{R}}| \alpha_{\tiem{R}}^{\|}H^{\|}_{\tiem{R}}m_{\tiem{R}}^{x(y)}
\label{eq:REtransverse}
\end{equation}
where the upper sign corresponds to the equation for $m_{\tiem{R}}^{x}$   and the lower sign for the $m_{\tiem{R}}^{y}$ one,
 $\Omega_{\tiem{R}}=zqm_{\tiem{R}}^{0}|\gamma_{\tiem{R}}| |J_{\tiem{TR}}| /\mu_{\tiem{R}}$ and $H^{\|}_{\tiem{R}}$ is the RE longitudinal field.
Equation \eqref{eq:REtransverse} shows that the perpendicular motion of the TM triggers the corresponding precessional motion of
the RE via the angular momentum transfer
(the first two terms of Eq. \eqref{eq:LLBT}, \emph{i.e.} via perpendicular components) with the same frequency
$\Omega_{\tiem{T}}$, but different amplitude, see Fig. \ref{schematics}(a).
During this dynamical process in some time interval the RE and TM magnetization have both the same sign of the $z$-component,
forming the transient ferromagnetic-like state seen experimentally \cite{Radu2011}.
Note that the subsequent precession has a frequency which is proportional to the exchange field and thus is extremely fast.
The motion of the TM around RE direction and vice versa occurs during and after the ferromagnetic-like state until
the system has relaxed to equilibrium.

\begin{figure}[tb!]
\includegraphics[width=8.0cm, trim=10 10 10 30]{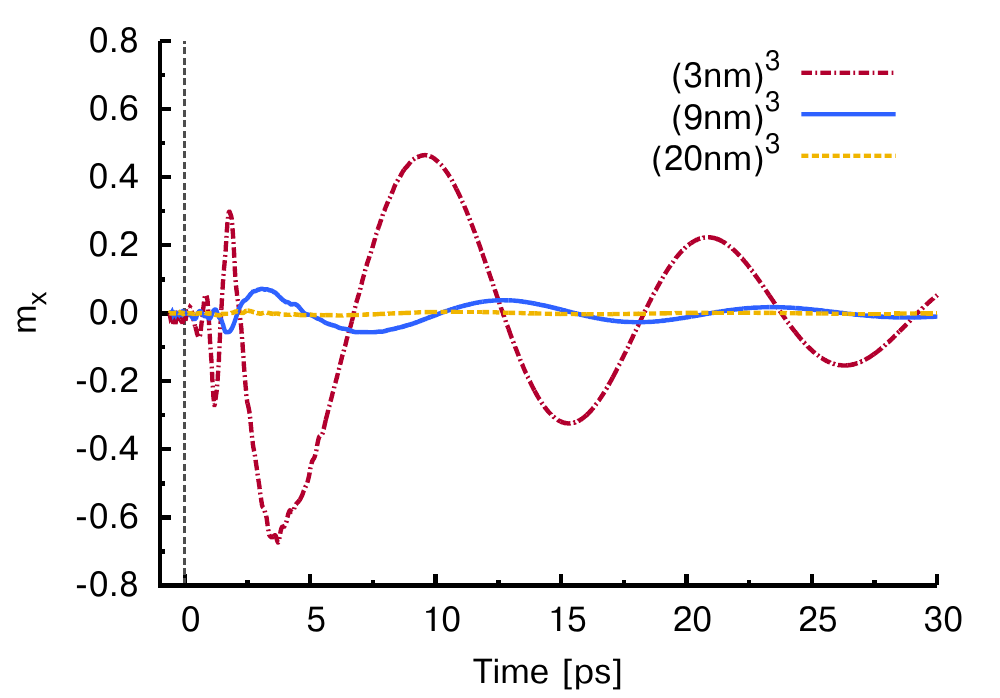}
\caption{Atomistic modeling of the system size dependence of the transverse magnetization components of the TM under ultrafast switching, showing cancelation of
the localized transverse magnetization components arising from exchange precession for larger system sizes. The time $t=0$ corresponds to
the end of the laser pulse.}
\label{size}
\end{figure}

An outstanding question is whether the magnetization precession, a central part of the process, can be observed experimentally
on a macroscopic sample.
We should recall that  in non-equilibrium at high temperatures the correlation between atomic sites is weak,
thus we cannot expect the precession  to occur with the same phase in the whole sample;
 an effect which would make the precession macroscopically unobservable.
To demonstrate the effect we present in Fig.~\ref{size}
the results of  atomistic switching simulations in GdFeCo for different
 system sizes ($T_{\text{max}}=2000$ K). In Fig. \ref{size}
we observe that for small system sizes
transverse oscillations with the frequency of an exchange mode are  visible,
consistent with the prediction of our analytical model.
However, in large system sizes of the order of $($20 nm$)^3$ it is averaged out, consistent with the excitation of localized exchange
modes with random phase. Note that the same effect happens for very high temperatures where the observed magnetization trajectory appears
close to linear; although we stress again the importance of a small perpendicular component to initiate the magnetization reversal,
which will occur on a local level as demonstrated by Fig.~\ref{size}.

In conclusion, the LLB equation
for a ferrimagnet describes the mutual relaxation of sublattices which occurs simultaneously under internal damping and
inter-sublattice exchange. This model allows us to present a simple picture of the magnetization reversal of GdFeCo in response to an
ultrafast heat pulse alone. The physical origin of this effect is revealed within the LLB equation as a dynamical reversal path resulting
from the instability of the linear motion. To trigger the reversal path a small perpendicular component is necessary. In practice this will
arise from random fluctuations of the magnetization at elevated temperatures. The perpendicular component grows in time resulting in ultrafast
magnetization precession in the inter-sublattice exchange field, also observed in atomistic simulations for small system sizes.
The switching is initiated by the TM which arrives at zero magnetization faster than the RE and responds dynamically to its exchange field.
Thus, the non-equivalence of the two sub-lattices is an essential part of the process. Switching into the transient ferromagnetic state occurs due
to large-amplitude precessional motion of the TM in the exchange field from the  RE and a slow dynamics of RE.

This work was supported by the European Community's Seventh Framework Programme (FP7/2007-2013) under grant agreements
NMP3-SL-2008-214469
(UltraMagnetron)  N 214810 (FANTOMAS), NNP3-SL-2012-281043 (FEMTOSPIN) and the Spanish
Ministry of Science and Innovation under the grant
FIS2010-20979-C02-02.


\end{document}